\documentclass[aps,prl,twocolumn]{revtex4}
\usepackage{float}
\usepackage{hyperref}
\usepackage{graphicx}
\usepackage[utf8]{inputenc}
\DeclareGraphicsExtensions{.png,.jpg,.eps,.pdf}
\usepackage{amsmath}
\usepackage{comment}
\usepackage{bm}
\usepackage{dsfont}
\usepackage{amsfonts}
\usepackage{xcolor}

\begin{document}

\title{Certifying entanglement of  spins on surfaces  using ESR-STM}

\author{Y. del Castillo$^{1,2}$, J. Fern\'{a}ndez-Rossier$^1$\footnote{On permanent leave from Departamento de F\'{i}sica Aplicada, Universidad de Alicante, 03690 San Vicente del Raspeig, Spain}$^,$\footnote{joaquin.fernandez-rossier@inl.int} }

\affiliation{$^1$International Iberian Nanotechnology Laboratory (INL), Av. Mestre Jos\'{e} Veiga, 4715-330 Braga, Portugal }
\affiliation{$^2$Centro de F\'{i}sica das Universidades do Minho e do Porto, Universidade do Minho, Campus de Gualtar, 4710-057 Braga, Portugal }

\date{\today}


\begin{abstract}

We propose a protocol to certify the presence of entanglement in  artificial on-surface atomic and molecular spin arrays using  electron spin resonance carried by scanning tunnel microscopes (ESR-STM). We first generalize the theorem that relates global spin susceptibility  as  an entanglement witness to the case of anisotropic Zeeman interactions, relevant for surfaces. We then propose a method to measure the spin susceptibilities of surface-spin arrays  combining  ESR-STM with atomic manipulation. Our calculations
show that entanglement can be certified in antiferromagnetically coupled spin dimers and trimers  with state of the art ESR-STM magnetometry.

\end{abstract}

\maketitle

Very much like electricity and magnetism, quantum entanglement is a naturally occurring phenomena. Entanglement lies at the heart of the most intriguing aspects of quantum mechanics\cite{bell04},  such as quantum teleportation\cite{bennett93,bouwmeester97,boschi98},  non-locality and  the emergence of exotic quantum states of matter\cite{kitaev06}. The notion that quantum entanglement is also a resource that can be exploited is at the cornerstone of the fields of quantum computing\cite{nielsen02}, quantum sensing\cite{degen17} and quantum communications\cite{wootters98}.  In order to  harness quantum entanglement,  it seems imperative  to develop tools to probe it\cite{friis19}.  This can be particularly challenging at the nanoscale\cite{heinrich2021}.

Quantum entanglement is predicted to  occur spontaneously in interacting spin systems\cite{arnesen01,amico09}, going from small clusters to crystals.  In the last few decades,  tools to create  and probe artificial spin arrays based on both magnetic atoms and molecules on surface have been developed\cite{hirjibehedin06,khajetoorians19,choi19}. There is now a growing interest in the exploitation of  quantum behaviour of this class of systems\cite{delgado17,chen22}.   Surface-spin artificial arrays can display exotic quantum magnetism phenomena, including spin fractionalization in  $S=1$ Haldane chains\cite{mishra21}, resonant valence bond states in spin plaquetes\cite{yang2021} and  quantum criticality\cite{toskovic16} that are known to host  entangled states.   However, experimental protocols that unambiguously determine the presence of entanglement in on-surface spins  are missing.

Here we propose a protocol to certify the presence of entanglement on artificial arrays of surface spins.  The approach relies on two ideas. First,  it was both proposed\cite{wiesniak05,brukner06,vedral08} and demonstrated experimentally\cite{brukner06,sahling15} in bulk systems, that the  trace of the spin susceptibility matrix can be used as an entanglement witness. Second, the development\cite{baumann15} of single-atom electron spin resonance using scanning tunneling microscopy (ESR-STM) that, combined with atomic manipulation,   has made it possible to carry out  absolute magnetometry\cite{natterer17,choi17} of surface spins. As illustrated in the scheme of Figure 1,  the magnetic field created by the different quantum states of the  surface spins can be measured by an ESR-STM active atomic spin sensor placed nearby. The  accurate determination of the  height of the corresponding peaks and their shift makes it possible to pull out both their occupation and magnetic moment, which permits one to determine the spin susceptibility. 

The global spin susceptibility is defined as the linear coefficient that relates the external magnetic field to the average  total  magnetization
\begin{equation}
\langle M_{\alpha}\rangle= \chi_{\alpha} B_{\alpha}
\label{susc0}
\end{equation}
where  
$\langle\rangle$ stands for statistical average in thermal equilibrium, and
\begin{equation}
M_{\alpha}= \sum_{i} m_{\alpha}^i=-\mu_B \sum_i g_{\alpha}S_{\alpha}^i
\label{MTOT}
\end{equation} 
where $\alpha=x,y,z$ labels the principal axis that diagonalize the susceptibility tensor and $i$ labels the site in a given spin lattice.

\begin{figure*}[t]
\centering
\includegraphics[width=0.85\textwidth]{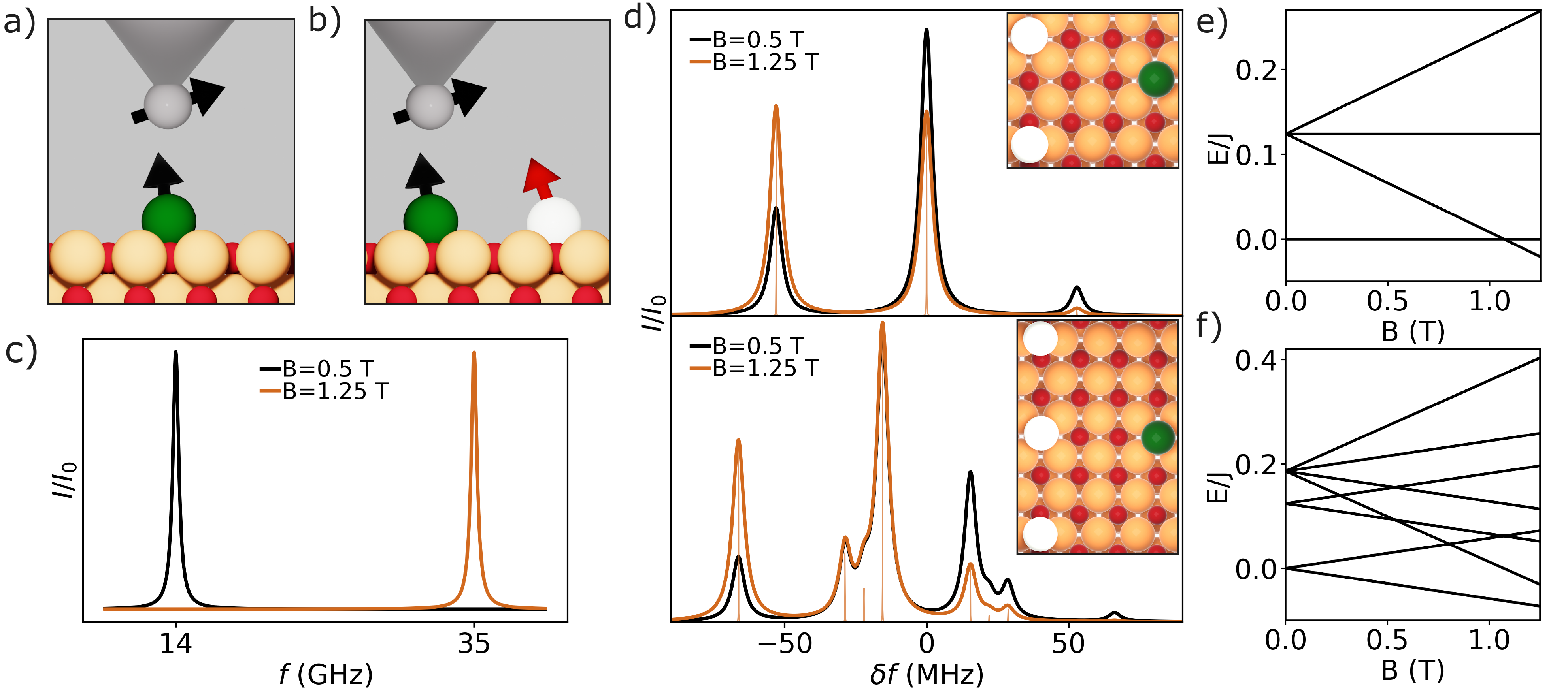}
\caption{Scheme of the proposed STM-ESR experimental protocol. (a) Characterization of the bare resonance  frequency $f_0$ of the  sensor.   (b) Determination of the $g$ coefficients  of a single unit (atom) of the on-surface spin array. (c)  ESR-STM readout of the bare sensor at two different fields.  (d) ESR-STM readouts of a sensor nearby antiferromagnetically coupled  dimer (top) and a trimer (bottom), with $J=124\mu$eV and $T=1$ K,  using different magnetic fields that lead to different occupations of the states. (e) and (f)  Energy spectra of the dimer and trimer as a function of $B$. }
\label{fig:setup}
\end{figure*}

The interaction between the spins of interest and the external field is given by:
${\cal H}_1=-\sum_{\alpha}  M_{\alpha}B_{\alpha}$.
For spin-rotational invariant Hamiltonians that commute with the Zeeman operator,  $[{\cal H}_0,{\cal H}_1]=0$, 
  the  spin susceptibility and the statistical variance
 of the total magnetization, $\Delta M_{\alpha}^2$ are related by
 \cite{wiesniak05}\footnote{See supplementary material for a derivation}:
\begin{equation}
\chi_{\alpha}= \beta\Delta M_{\alpha}^2=\beta \left(\sum_{i,j}\langle m_{\alpha}^i m_{\alpha}^j\rangle -(\sum_i \langle m_{\alpha}^i\rangle )^2 \right)
\label{flures}
\end{equation}
where $\beta=\frac{1}{k_B T}$ .   The variance for an isolated spin satisfies 
$ \sum_{\alpha} g_{\alpha}^2 \Delta S_{\alpha}^2\geq g_{\rm min}^2 S$
where $g_{\rm min}={\rm Min}(g_x,g_y,g_z)$.
For a multi-spin  system it has been shown\cite{hofmann03} that, {\em for non-entangled states,  the variance of total magnetization is at least as large as the sum over variances of individual magnetizations}. We thus can write the equation:
\begin{eqnarray}
{\rm Tr}\,\chi=
\sum_{\alpha}\chi_{\alpha} =\chi_x+\chi_y+\chi_z=\nonumber\\
=\beta \sum_{\alpha}\left(\sum_{i,j}\langle m_{\alpha}^i m_{\alpha}^j\rangle -(\sum_i \langle m_{\alpha}^i\rangle )^2 \right)
\geq \frac{g_{\rm min}^2\mu_B^2S}{k_BT}N
\label{EW}
\end{eqnarray}
where $N$ is the number of spins and $S$ is the spin quantum number ($S=\frac{1}{2},1,...$).
Equation (\ref{EW}) generalizes the result derived by Wiesnak {\em et al.}\cite{wiesniak05} to the case of anisotropic $g$ factor,
relevant for surface spins\cite{ferron19}, and constitutes the starting point to our proposed entanglement certification method.  If the spin-susceptibilities are measured, and violate the inequality of eq. (\ref{EW}), the presence of entanglement is certified. Equation (\ref{EW}) is a {\em sufficient} condition for entanglement, but not necessary: entanglement may be present, but go un-noticed, if the susceptibility satisfies eq.  (\ref{EW}).

We discuss now  how  to  certify entanglement   using  a single STM-ESR-active atom, that will be used as a sensor, placed nearby the spin array of interest.  Specifically, we consider the case of   dimers\cite{yang17,kot22} and  trimers\cite{yang2021} of antiferromagnetically coupled $S=1/2$  spins
on MgO\cite{yang2021}, the canonical surface for STM-ESR. The choice of $S=1/2$ rules out single ion anisotropies. In the suppl. material we show how the role of  dipolar interactions can be neglected in most cases. 
Examples of $S=1/2$   adsorbates deposited on MgO include  TiH \cite{yang17,yang18,yang19,seifert20,steinbrecher21,veldman21,kot22}, Cu\cite{yang18}, dimers of alkali atoms\cite{kovarik22} and even organic molecules\cite{lutz22}.

The spectra of the dimer and trimer are shown in Figure 1(e,f) as a function of the applied field. The dimer shows a singlet  ground state with wave function $\frac{1}{\sqrt{2}}\left(|\uparrow\downarrow\rangle -|\downarrow\uparrow\rangle\right)$ and a triplet of excited states $(S=1,S_z)$. The $S=1,\,S_z=\pm 1$ are  product states, whereas the $S=1,\,S_z=0$ is entangled.  The trimer ground state is a $S=1/2$ doublet with wave functions:
\begin{equation}
|\frac{1}{2},+\frac{1}{2}\rangle=
 -\frac{1}{\sqrt{6}}
 \left(|\uparrow\uparrow\downarrow\rangle+
 |\downarrow\uparrow\uparrow\rangle\right)
 +
 \sqrt{\frac{2}{3}}
 |\uparrow\downarrow\uparrow\rangle
\end{equation}
and an analogous formula for $|S=\frac{1}{2},S_z=-\frac{1}{2}\rangle$. Therefore, both structures, dimer and trimer, have entangled ground states.  The lowest energy excited state, with excitation energy $J$, is also a $S=1/2$ doublet,  whereas the highest energy multiplet has $S=3/2$ and excitation energy $\frac{3J}{2}$.

Experiments\cite{choi17} show that the ESR spectrum, i.e., the $I_{DC}(f)$  curve  of the  sensor adatom, under the influence of a nearby  spin-array,  can be described by:
     \begin{eqnarray}
     I_{DC}(f)= \sum_n p_n L(f-f_n)
     \label{eq:ESR-STM}
     \end{eqnarray}
     where the sum runs over the eigenstates  of the Hamiltonian of the spin array, $\left({\cal H}_0+{\cal H}_1\right)|n\rangle=E_n|n\rangle$, $L$ is a lorentzian curve centered around the resonant frequency $f=f_n$, given by the expression 
     \begin{equation}
    f_n(\vec{r})=\frac{\mu_B}{h} \sqrt{\sum_{\alpha} \left[ g_{\alpha}^{\rm sensor}
    (B_{\alpha}+b_{n,\alpha}(\vec{r}))\right]^2}
    \label{wsen}
    \end{equation}
Here, $h=2\pi \hbar$, $g_{\alpha}^{\rm sensor}$   are the components of  the gyromagnetic tensor of the sensor, $B_{\alpha}$ represents the the external field and $b_{n,\alpha}(\vec{r})$
is the $\alpha$ component of the  the stray field generated by the spin-array in the state $n$. The vector is given by
    \begin{equation}
    \vec{b}_n(\vec{r})= \frac{\mu_0}{4\pi}\sum_{i=1,N} \frac{3 (\vec{m}^i(n)\cdot\vec{d}^i)\vec{d}^i}{(d^i)^5} -\frac{\vec{m}^i(n)}{(d^i)^3}
    \label{stray}
    \end{equation}
    where $\vec{r}^i$ are the positions of the surface spins and $\vec{r}$ is the position of the sensor, $\vec{d}^i=\vec{r}-\vec{r}^i$ and $\vec{m}^i(n)$  is a vector given by with components by
    \begin{equation}
        m_{\alpha}^i(n)= -g_{\alpha}\mu_B \langle n|S_{\alpha}i|n\rangle
    \end{equation}
    the encode the quantum-average magnetization of the atomic moments of the spin array for state $n$. Equation \ref{eq:ESR-STM} implicitly assumes the absence of resonant spin-flip interactions between the sensor spin and the spin-array.

   In Figure \ref{fig:setup}d we show the spectra for a sensor nearby a dimer and a trimer of antiferromagnetically coupled spins, in the presence of a external field along the $z$ direction, assuming a broadening $\Delta f=5\, \rm{MHz}$ \cite{baumann15}.
The frequency shift of  the $n$ peak, with respect to the bare sensor, are governed by the stray field $\vec{b}_n(\vec{r})$ generated by that state.  
    For the dimer, the ESR-spectrum shows a prominent peak, that has contributions from both the $S=0,\,S_z=0$  and $S=1,\,S_z=0$ states, whose stray field vanish. As a result, the prominent peak has the frequency of the bare sensor. The two smaller peaks, symmetrically located around the central peak,  correspond to the $S=1, S_z=\pm 1$ states. In the case of the trimer up to eight different peaks appear in our simulation, as the different doublets generate states with  different magnetization and stray field.

     We now express the average magnetization and the susceptibility in terms of the occupations $p_n$  and the magnetic moment of each state, the two quantities that can be obtained from ESR-STM. For spin-rotational invariant systems, we  write  eq. (\ref{susc0}) as:
    \begin{equation}
     \langle M_{\alpha}\rangle=\sum_n p_n M_\alpha(n)   
    \end{equation}
    where $p_n= \frac{1}{Z} e^{-\beta E_n}$ is the equilibrium occupation of the $n$ eigenstate and $M_\alpha(n)=-g_{\alpha}\mu_B S_\alpha(n)$, and $S_\alpha(n)$ is the (half-)integer projection of the total spin operator of the spin-array along the $\alpha$ direction. 
    The susceptibility would be given by 
    \begin{equation}
    \chi_{\alpha}=\sum_n \frac{dp_n}{dB_\alpha} M_\alpha(n).    
    \end{equation}

     The experimental protocol to determine the $p_n$ is the following\cite{choi17}. For an ESR spectrum with $N_p$ visible peaks, we obtain the ratios  $r_n=\frac{I_n}{I_0}$ where $I_0$ corresponds to the largest peak that arises from a single state.  For trimers,  $I_0$ is the peak that arises from the ground state. For dimers, the ground state peak has also contributions from the $S=1,S_z=0$ state and therefore we  normalize with respect to the  $S=1,S_z=-1$ peak.  It  has been demonstrated  \cite{choi17} that the ratio of thermal occupations is the same than the ratio of Boltzman factors $r_n=\frac{p_n}{p_0}=\frac{I_n}{I_0}$.   Assuming that the peaks exhaust the occupations of the states, we  obtain $p_n$ out of the ratios  (see suppl. mat.) 
      
      The determination of the $M_\alpha(n)$ can be done in two ways. First, 
      the (half-)integer values of $S_{\alpha}$ could be assigned  by inspection of the ESR spectrum relying on  the fact that the states with largest $S_{\alpha}$ have larger stray field. We have verified\cite{del23} the feasibility of a second approach where the readout of the ESR spectra is repeated with the sensor in different locations, mapping the stray field of every state in real space, and pulling out the atomic magnetization averages, inferring  $M_{\alpha}(n)$ thereby.

The protocol to measure the spin susceptibility of a spin array has the following steps, that have to be implemented for the
three orientations $\alpha=x,y,z$  of the external field:
1. The bare resonant frequencies for the sensor
$f_{0\alpha}=\frac{\mu_B}{h} g_{\alpha}^{\rm sensor} B_{\alpha}$ (Figure 1a)
are determined in the {\em absence} of the spin array.

2. The $g_{\rm min}$ for the individual spins that form the array are determined. If they are ESR-STM active, a conventional resonance experiment is carried out, for the three directions of the field. Otherwise, the  nearby ESR-STM sensor atom can be used \cite{choi17}.
      
3. The $M_{\alpha}(n)$ and $p_n$ are obtained as explained above, for two values of $B_{\alpha}^{(1)}$ and $B_{\alpha}^{(2)}$.  This permits us to obtain 
$\langle M_{\alpha} (B_{\alpha}^{(1)})\rangle$
and
$\langle M_{\alpha} (B_{\alpha}^{(2)})\rangle$
for both fields and $\chi_{\alpha}=\frac{\langle M_{\alpha} (B_{\alpha}^{(2)})\rangle-\langle M_{\alpha}(B_{\alpha}^{(1)})\rangle}{B_{\alpha}^{(2)}-B_{\alpha}^{(1)}}$
        and compare to eq. (\ref{EW}) to establish the presence of entanglement.

In the top panels of figure 2 we show the spin susceptibility for dimers and trimers as a function of temperature, together with the EW boundary, assuming isotropic  factors, $g_{\alpha}=2$. We define the temperature $T^*$, below which the trace of the susceptibility matrix is smaller than the entanglement witness limit, so that equation (\ref{EW}) is not satisfied,  and therefore the presence of entanglement is certified.  The $\rm{Tr} \,\chi(T)$ curves are very different for dimer and trimer.  At low T, the dimer is in the singlet state and its susceptibility vanish. In contrast, the trimer 
with $J>> k_B T$,  the excited states play no role and the spin susceptibility is governed by the two states of the $S=1/2$ ground state doublet.

\begin{figure}[ht]
\centering
\includegraphics[width=0.84\linewidth]{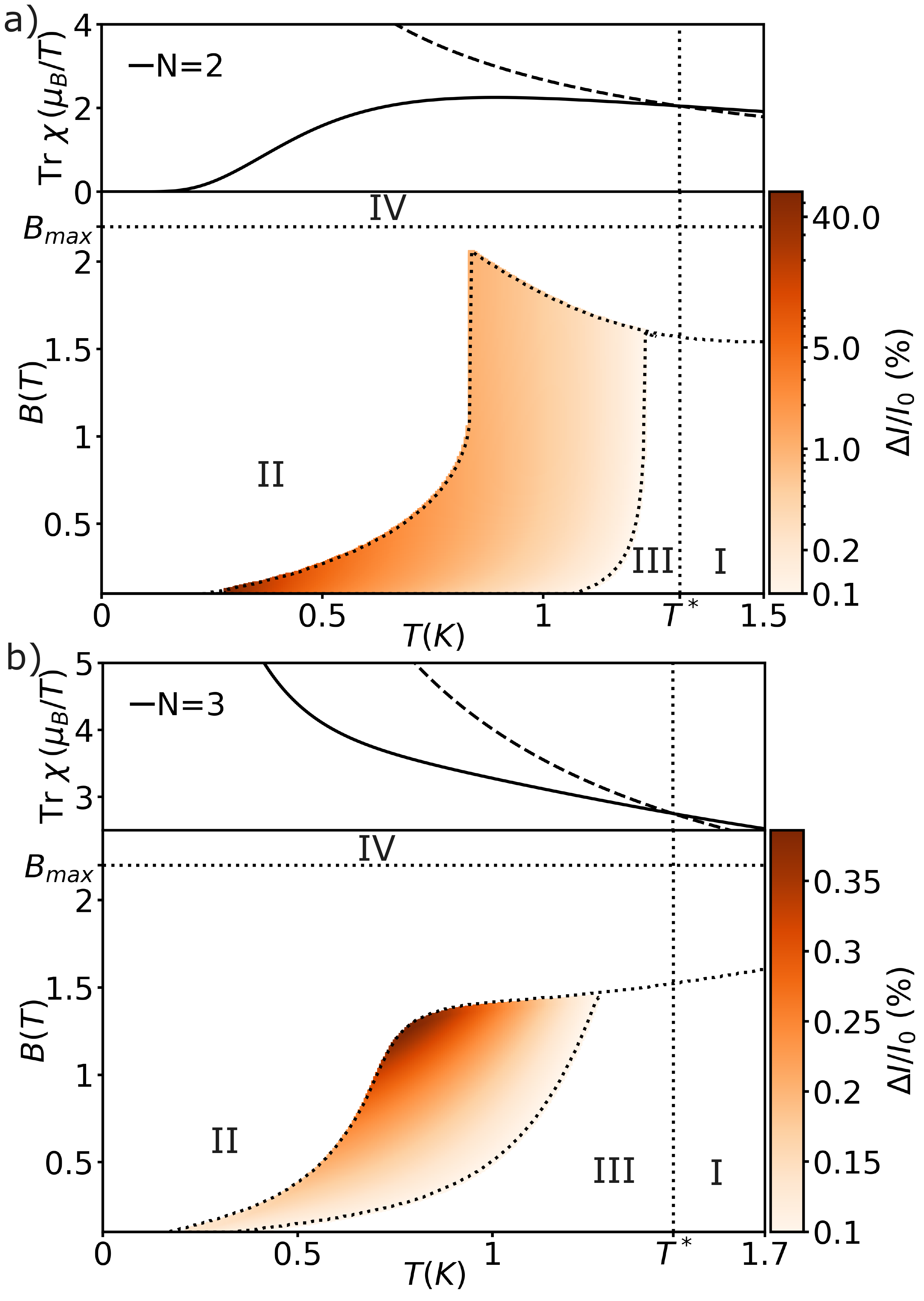}
\caption{ Entanglement certification window (see text) for a dimer (a) and trimer (b) with with $J= 124\mu$eV, using  a sensor  located 0.9 nm away. Top panels show $\rm{Tr}\,\chi(T)$ and the EW condition of eq. (\ref{EW}). Color map inside the ECWs represents the largest experimental error $\Delta I/I_0$ that still makes entanglement certification possible, assuming $\eta=5$ in eq. \ref{deltaI}. }
\label{fig:Frequency}
\end{figure}

The implementation of the entanglement certification protocol will only be possible if the stray fields induce a shift larger than the spectral resolution of ESR-STM.
For ESR-STM , $\Delta f\simeq 4 $MHz\cite{baumann15}, determined by the $T_1,T_2$ times of the adsorbate in that system. In the simulations of Figure 1 we have used $\Delta f= 5MHz$.  In addition, the experiment  has to meet four  experimental conditions. 
Condition $I$:  temperature  $T$ has to be smaller than $T^*$. Condition $II$: $M_{\alpha}(B_{\alpha})$  has to be linear. We define $B^*$ as the upper value of the field above which $\frac{d M_{\alpha}}{dB_{\alpha}}$ deviates from its value for $B=0$ more than 5 percent.

Condition $III$: the relative error in the current of the reference peak  $\frac{\Delta I}{I_0}$  should  be small enough to allow for the accurate entanglement certification.  Specifically,  the error in the spin susceptibility associated to
$\frac{\Delta I}{I_0}$, should be  smaller, by a factor $\eta>>1$,   than the difference between $\chi_{\alpha}$ and the EW condition. In the suppl. mat. show this condition is 
\begin{equation}
   \frac{\Delta I}{I_0}   
< \frac{B |{\rm Tr\,} \chi-\chi_{EW}|}{ 6 \eta p_0 \sum_n \left(\left( 2+ r_n + (N_p-2)p_0 \right)
    |M_z(n)| \right) }.
     \label{deltaI}
\end{equation}
Condition $IV$: the external field must not  drive the sensor resonant frequency above the excitation bandwidth of ESR-STM.  The record, so far, 
is  $f_0 = 61 GHz$ \cite{kot22}, that translates into  $B\simeq2.2$ T  for $S=1/2$  Ti-H adsorbates.  

The independent satisfaction of  conditions $I$ to $IV$ occurs only in a region of  the $(B,T)$ plane, defining four boundary lines, one per condition. We define the {\em entanglement certification window (ECW)} as the area where the four conditions are satisfied simultaneously.
In Figure \ref{fig:Frequency} we show the ECW, assuming $\eta=5$ 
and imposing for condition $III$ that $\Delta I/I_0$ can not go below $10^{-3}$, for the case of a dimer (a) and a trimer (b) with  the same $J$ reported by Yang {\em et al.}\cite{yang2021}  $J=124\mu eV$ (30 GHz). The colour code signals the maximal $\Delta I/I_0$ that satisfies eq. (\ref{deltaI}).

\begin{figure}[h]
\centering
\includegraphics[width=1\linewidth]{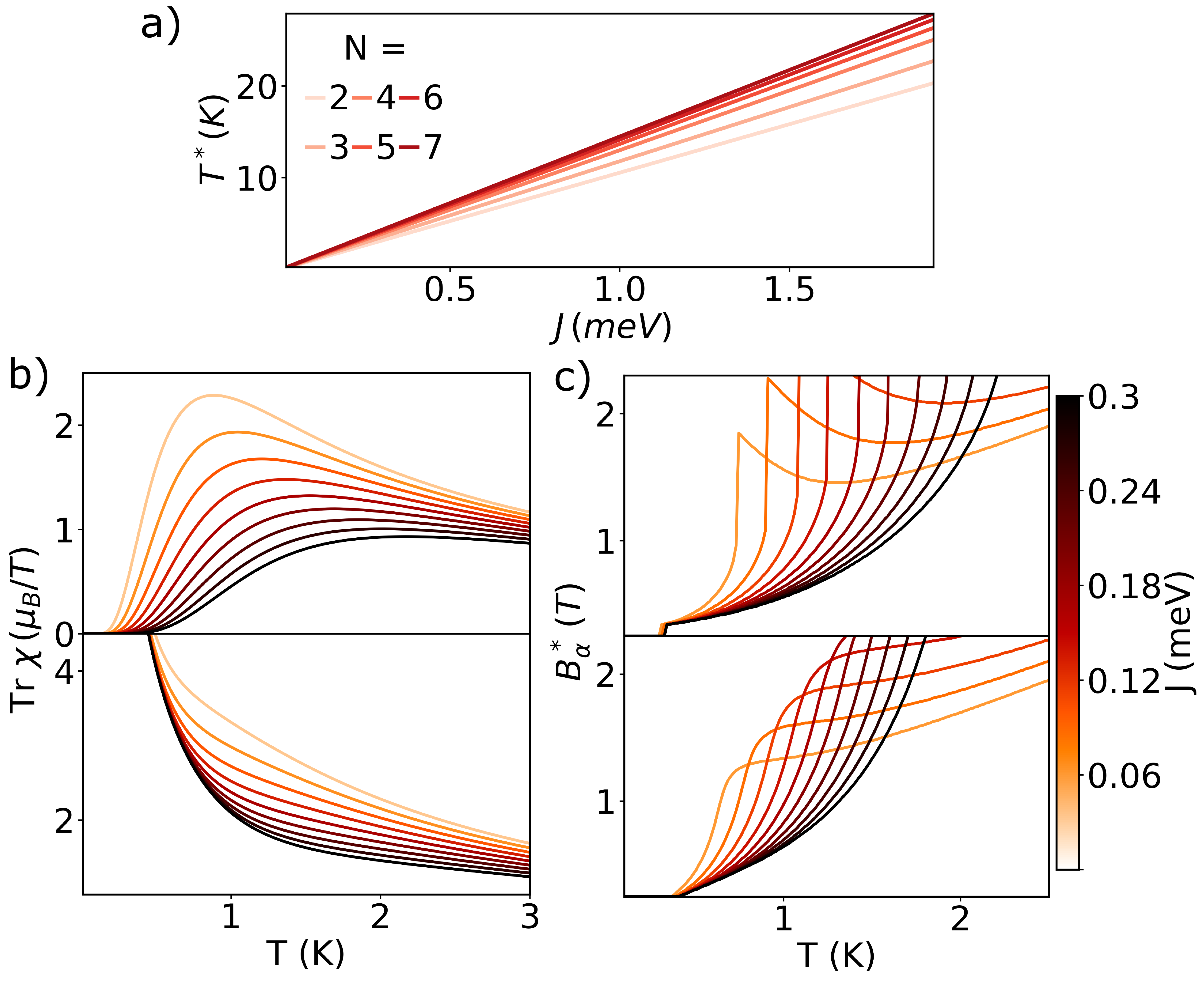}
\caption{(a) $T^*(J)$ for different chain sizes.(b,c) Temperature and $J$ dependence of the spin susceptibility (b) and $B^*$ (c) of dimer (top) and trimer (bottom).  }
\label{fig:Limits}
\end{figure}

We now discuss the effect of the value of $J$ on the size of the ECW. In Figure 3a  we show that  $T^*$ is strictly linear with $J$ for spin chains with $N$ sites,  and the slope is an increasing function of the number of $N$.  Therefore, both long chains and large $J$ increase the crossover temperature. As of condition two, our numerical simulations show that $B^*$ is an increasing function of both $T$ and $J$: the magnetization vs field curves remain linear as long as the Zeeman energy is smaller than the other two relevant energy scales $J$ and $T$. Condition three is governed both  by the ratio  $B/p_0(B,T)$ (see eq. \ref{deltaI} ) and the susceptibility.
In Figure 3b we show the $J$ dependence of the $\chi(T)$ curves. We see that reducing $J$ tends to increase $\chi$, at a fixed value of $T$. Again, this reflects the competition between exchange and Zeeman energies. Therefore, increasing $J$ pushes $T^*$ up and pushes the $\chi(T)$ away from the EW boundary.  The only advantage of smaller values of  $J/K_BT $ is the presence of more peaks in the $I(f)$ spectra that makes the spectra clearly distinguishable from simpler structures.

In conclusion, we propose a method to determine the presence of quantum entanglement in on-surface spin arrays using ESR-STM.
For that matter, the trace of the spin susceptibility of the spin arrays has to be measured taking advantage of  the STM potential 
{\em i)} to manipulate atoms and build spin structures, and {\em ii)} to simultaneously resolve the stray fields  and occupation probabilities of several quantum states of nearby spins\cite{choi17}. Our calculations show that the experiments can be carried out with the state of the art ESR-STM instrumentation in terms of both  spectral resolution ($\Delta f$) and relative error in the current  $\Delta I/I_0$\cite{choi17}.   The use of STM to certify the presence of entanglement would thus add a new functionality to this versatile tool and would  open a new venue in quantum nanoscience\cite{heinrich2021}. 
 
We acknowledge Antonio Costa for useful discussions. 
J.F.R.  acknowledges financial support from 
 FCT (Grant No. PTDC/FIS-MAC/2045/2021),
 SNF Sinergia (Grant Pimag),
FEDER /Junta de Andaluc\'ia, 
(Grant No. P18-FR-4834), 
Generalitat Valenciana funding Prometeo2021/017
and MFA/2022/045,
and
 funding from
MICIIN-Spain (Grant No. PID2019-109539GB-C41).
%
%
%
%
YDC acknowledges funding from FCT, QPI,  (Grant No. SFRH/BD/151311/2021) and thanks the hospitality of the Departamento de F\'isica Aplicada at the Universidad de Alicante.



\bibliographystyle{apsrev4-1}
\bibliography{biblio}{}

\end{document}


\title{Certifying entanglement of  spins on surfaces  using ESR-STM:
Supplemental material}

\author{Y. del Castillo$^{1,2}$, J. Fern\'{a}ndez-Rossier$^1$\footnote{On permanent leave from Departamento de F\'{i}sica Aplicada, Universidad de Alicante, 03690 San Vicente del Raspeig, Spain}$^,$\footnote{joaquin.fernandez-rossier@inl.int} }

\affiliation{$^1$International Iberian Nanotechnology Laboratory (INL), Av. Mestre Jos\'{e} Veiga, 4715-330 Braga, Portugal }
\affiliation{$^2$Centro de F\'{i}sica das Universidades do Minho e do Porto, Universidade do Minho, Campus de Gualtar, 4710-057 Braga, Portugal }

\date{\today}

\begin{appendix}
\maketitle

\section{Derivation of the spin-susceptibility in terms of the spin-fluctuation variance}
We demonstrate here equation 4 in the main text. 
We consider a spin Hamiltonian with two parts:
\begin{equation}
{\cal H}= {\cal H}_0 -\sum_{\alpha} M_{\alpha} B_{\alpha}
={\cal H}_0+ {\cal H}_1
\end{equation}
where ${\cal H}_0$ describes a lattice Hamiltonian of interacting spins, $B_{\alpha}$ is the value
of the $\alpha=x,y,z$  external magnetic field, and $M_{\alpha}$ is the total spin magnetization, defined in the text.
Importantly, we assume both that ${\cal H}_0$  {\em does not depend on $B$} and  $[H_0, M_{\alpha}]=0$.

 We now write  the susceptibility
 as the derivative of the average magnetization as a function of the applied field:
 \begin{equation}
 \chi_{\alpha}\equiv
 \frac{\partial \langle M_{\alpha}\rangle}{\partial B_{\alpha}}
 \label{susc0}
 \end{equation}
 where $<>$ stands for the statistical 
  expectation value, 
$ \langle O \rangle=\frac{1}{Z}Tr(e^{-\beta H} O)$
 where  
$ Z=Tr(e^{-\beta {\cal H}} )$
is the partition function
 and $\beta=\frac{1}{k_B T}$. 
 Equation (\ref{susc0}) is the derivative of a division that we express as $f/Z$, where
 $f= Tr(e^{-\beta {\cal H}} M_{\alpha})$. Therefore, we have:
 $ \chi_{\alpha}=\frac{f'Z-fZ'}{Z^2}$

 We thus need to compute two derivatives:
 \begin{equation}
 f'=\frac{\partial}{\partial B_j^{\beta}} Tr(e^{-\beta {\cal H}} M_{\alpha})
 \end{equation}
 and
 \begin{equation}
 Z'=\frac{\partial}{\partial B_j^{\beta}} Tr(e^{-\beta {\cal H}} )
 \end{equation}
 
 We now make two important remarks. First, 
 the values of the traces do not depend on the choice of the basis set. Therefore, in the following, we can always
 assume the only dependence of the traces on the magnetic field occurs only via the explicit dependence of $H$ on $B$, and not
 on the dependence of the quantum states that we need to use to compute the trace.   Therefore, the derivative can be taken inside the trace,
 $f'=Tr(\frac{\partial}{\partial B_j^{\beta}} e^{-\beta {\cal H}} M_{\alpha})$.
Since the operator $M_{\alpha}$ does not depend on $B$, and the $Tr(AB)=Tr(BA)$ we can write
  \begin{equation}
 f'=Tr(M_{\alpha}\frac{\partial}{\partial B_{\alpha}} e^{-\beta {\cal H}} )
 \end{equation}
 The second important remark is that, since we assume ${\cal H}_0$ commutes with $M_{\alpha}$, we can write the derivative with respect to $B_{\alpha}$ as
  \begin{equation}
 \frac{\partial}{\partial B_{\alpha}} e^{-\beta {\cal H}} = -\frac{1}{k_BT}  \frac{\partial {\cal H}_1}{\partial B_{\alpha}}e^{-\beta ({\cal H}_0+{\cal H}_1)} 
 \label{chain}
 \end{equation}

Using the identity  $\frac{\partial {\cal H}}{\partial B_{\alpha}}= -M_{\alpha}$ we obtain $\frac{\partial}{\partial B_{\alpha}} e^{-\beta {\cal H}} = \frac{1}{k_B T} M_{\alpha}e^{-\beta {\cal H}}$ and  we  arrive to 

\begin{equation}
f'=\frac{1}{k_BT}Tr(M_{\alpha}^2 e^{-\beta {\cal H}}  )
=\beta Z \langle  M_{\alpha}^2\rangle
\end{equation}
 
 We now compute $Z'$, using
 the same arguments than below
 and we write:
 \begin{equation}
\frac{\partial}{\partial B_{\alpha}} Tr(e^{-\beta {\cal H}} )=
Tr\frac{\partial}{\partial B_{\alpha}}e^{-\beta {\cal H}}
\end{equation} 
We now use eqs. \ref{chain} and we obtain:
 \begin{equation}
Z'=
\frac{1}{k_B T}Tr(M_{\alpha} e^{-\beta {\cal H}} )= \frac{Z}{k_B T} \langle M_{\alpha}\rangle
\end{equation} 

Thus, we arrive to :
 \begin{equation}
 \chi_{\alpha}=
 \frac{1}{k_B T} \left(\langle  M_{\alpha}^2\rangle -
 \langle  M_{\alpha}\rangle^2
   \right)
 \end{equation}
 We now write $M_{\alpha}=\sum_i m_{\alpha}^i$ and we obtain equation 4.

\section{Inferring $p_n$}
A given ESR-STM spectrum has a series of $N_p$ peaks that we label 
as $0$, $1$,... $N_p-1$.  We assume the height of each peak is proportional to the sum of the occupation probabilities of the states that contribute to that peak.
We distinguish two cases: when there is a one to one relation between peaks and states (\emph{unclustered})  and otherwise (\emph{clustered}). 

Eigenstates of the spin-rotational Hamiltonians considered here can be classified according to their total spin $S$ and the projection of their total spin along the direction of the field,  $S_{\alpha}$, that takes integer or half-integer values only. 
For the dimer and the trimer, we have found that the absolute value of the frequency shift is an increasing function of $S_{\alpha}$. This can be rationalized as follows: if we express the stray field as a multipolar expansion, the leading order term is governed by the average magnetization.  The differences of stray field for states with the same $S_\alpha$ can be rationalized in terms of their different quadrupolar spin moment.

All $S_z=0$ states have a vanishing stray field and contribute to the same peak, whose frequency is identical to the bare frequency. In some cases, states  with the same finite $S_z$ may have identical stray fields at some point, so that they contribute to the same peak in the ESR-STM spectrum. For the dimer and the trimer, we have never found that two peaks with different $S_\alpha$ have the same stray field, but this could happen if the sensor is too far away, or in some highly symmetric sensor position.

We assign the label $0$ to  the largest peak 
that arises from an individual  state and we define the ratio between different peaks as:
\begin{equation}
    r_n=\frac{p_n}{p_0}
\label{rn}
\end{equation}
For unclustered peaks, $p_n$ is the probability occupation of an individual state.
For clustered peaks, instead, $p_n$ is a sum of probabilities of states, all with the same $S_{\alpha}$. Therefore, for the sake of the evaluation of the magnetization and susceptibility, we do not need to know the breakdown contribution of each state in a peak.

We further assume that the only states contributing to visible peaks have a finite probability and we write:
\begin{equation}1=\sum_{n=0,N_p-1}p_n=p_0+ \sum_{n=1,N_p-1} p_n 
\end{equation}
We divide this equation by $p_0$ and we invert, to obtain:
\begin{equation}
    p_0=\frac{1}{1+\sum_{n=1,N_p-1}r_n}
\label{p0}
\end{equation}
This permits us  to pull out $p_0$ and from the relative heights of the other peaks, $p_n$ with $n>0$.

\section{Estimation of the required  $\Delta I/I_0$}

The determination of the susceptibility lies on the capability to measure variations of peak heights in the ESR spectra, 
In the following, we assume that the minimal variation that can be detected is $\Delta I$. 
This sets the scale to the minimal variation for $p_n$, $\langle M_{\alpha}\rangle $ and $\chi_{\alpha}$. Here we obtain these relations.

\subsection{Estimation of uncertainty  on the probabilities}

Starting with equations (\ref{rn},\ref{p0}), the uncertainty in $r_n$ is given by:
\begin{equation}
\Delta r_n = \left(\left|\frac{dr_n}{dI_n}\right|+\left|\frac{dr_n}{dI_0}\right|\right)\Delta I 
= \left(\frac{1}{I_0}+ \frac{I_n}{I_0^2}\right)\Delta I= \frac{\Delta I}{I_0}\left(1+ r_n\right).
\end{equation}

We now infer the error on the estimation of $p_n= p_0 r_n$ using the formula: 
\begin{equation}
    \Delta p_n = p_0 \Delta r_n + \sum_{n'} |\frac{dp_0}{dr_{n}'}|\Delta r_{n'}
\end{equation}
where
$$
\left|\frac{dp_0}{dr_{n'}}\right|=\left|\frac{d}{dr_n}\frac{1}{1+\sum_{n'=1,N_p-1}r_{n'}}\right|
= \frac{1}{(1+\sum_{n'=1,N_p-1}r_{n'})^2}=p_0^2 .$$

We thus arrive to the following expression:
\begin{equation}
    \Delta p_n =  p_0 \Delta r_n +p_0^2 \sum_{n'\neq 0}  \Delta r_{n'}
\end{equation}
which can be written as
\begin{equation}
    \Delta p_n = p_0 \frac{\Delta I}{I_0}\left( \left(1+ r_n\right) +p_0 \sum_{n'\neq 0}  \left(1+ r_{n'}\right)\right).
\end{equation}

We now separate the sum, as so:
\begin{equation}
    \Delta p_n = p_0 \frac{\Delta I}{I_0}\left( \left(1+ r_n\right) + \sum_{n'\neq 0}p_0  + \sum_{n' \neq 0} p_0 r_{n'}\right).
\end{equation}
Then we use $1=\sum_n p_n$ and $p_0 r_{n'}= p_{n'}$ so that $\sum_{n'\neq 0} p_0 r_{n'}= \sum_{n'\neq 0} p_{n'}=1-p_0$:

\begin{equation}
    \Delta p_n = p_0 \frac{\Delta I}{I_0}\left( \left(1+ r_n\right) + (N_p-1)p_0  + (1-p_0) \right).
\end{equation}
Therefore

\begin{equation}
    \Delta p_n = p_0 \frac{\Delta I}{I_0}\left( 2+ r_n + (N_p-2)p_0 \right).
\end{equation}


\subsection{Estimation of error of $\langle M_\alpha \rangle$ and $\chi$}
We use eq. (2) and estimate its error as:
\begin{equation}
    \Delta M= \sum_n \left|\frac{d\langle M_{\alpha} \rangle}{dp_n}\right|\Delta p_n
\end{equation}

Then, we use the results above to write:
\begin{equation}
    \Delta \langle M_{\alpha}\rangle=  p_0\frac{\Delta I}{I_0} \sum_n (\left( 2+ r_n + (N_p-2)p_0 \right) |M_{\alpha}(n)| ).
    \end{equation}
    
Now, we estimate the error in the susceptibility as:
    
    \begin{equation}
        \Delta \chi_{\alpha}=\frac{2\Delta M_{\alpha}}{\Delta B} =\frac{2p_0\frac{\Delta I}{I_0} \sum_n (\left( 2+ r_n + (N_p-2)p_0 \right) |M_{\alpha}(n) |)}{\Delta B}>\frac{2p_0\frac{\Delta I}{I_0} \sum_n (\left( 2+ r_n + (N_p-2)p_0 \right) |M_{\alpha}(n) |)}{ B}
    \end{equation}
    
\noindent and the error in the trace of the susceptibility is:
    \begin{equation}
        \Delta ({\rm Tr} \chi)= \sum_{\alpha} \Delta \chi_{\alpha}
        =\sum_{\alpha} \frac{2p_0\frac{\Delta I}{I_0} \sum_n (\left( 2+ r_n + (N_p-2)p_0 \right) |M_{\alpha}(n) |)}{\Delta B}
        \geq\sum_{\alpha}\frac{2p_0\frac{\Delta I}{I_0} \sum_n (\left( 2+ r_n + (N_p-2)p_0 \right) |M_{\alpha}(n) |)}{ B}.
    \end{equation}
    
For the symmetric g-factor case that we have implemented in the manuscript we simply write:
    \begin{equation}
        \Delta ({\rm Tr} \chi)
        \geq 3\frac{2p_0\frac{\Delta I}{I_0} \sum_n (\left( 2+ r_n + (N_p-2)p_0 \right) |M_{\alpha}(n) |)}{ B}.
        \label{eq:DeltaChii}
    \end{equation}

From here we obtain the condition that relates $\Delta I/I_0$ to $\Delta Tr\chi$ which requires that:
\begin{equation}
        \frac{B}{6p_0\frac{\Delta I}{I_0} \sum_n (\left( 2+ r_n + (N_p-2)p_0 \right) |M_{\alpha}(n) |)}\Delta ({\rm Tr} \chi)
        \geq \frac{\Delta I}{I_0}.
        \label{eq:DeltaChi}
    \end{equation}
    
We now require that the error in the susceptibility is much smaller than its difference with the EW limit:
\begin{equation}
     \Delta(Tr  \chi)< \frac{|Tr( \chi)-\chi_{EW}|}{\eta}
     \label{eq:impose}
\end{equation}       
with $\eta>>1$.

We put together equations (\ref{eq:DeltaChi},\ref{eq:impose}) to obtain:
\begin{equation}
\frac{\Delta I}{I_0}
     \leq \frac{|Tr( \chi)-\chi_{EW}|}{\eta}\frac{B}{6p_0\frac{\Delta I}{I_0} \sum_n (\left( 2+ r_n + (N_p-2)p_0 \right) |M_{\alpha}(n) |)}.
\end{equation}

\section{ Spin susceptibility for an antiferromagnetic dimer.}
We derive the expression of the susceptibility for an AF spin dimer.
The analytical spin susceptibility at zero external field is given by the following expression:

\begin{equation}
\chi_{\alpha}= \beta\Delta M_{\alpha}^2=\beta \left(\sum_{i,j}\langle M_{\alpha}^i M_{\alpha}^j\rangle -(\sum_i \langle m_i^{\alpha}\rangle )^2 \right).
\label{chi}
\end{equation}

For a Heisenberg dimer, we have a singlet and a triplet. The energies for each are $ E_{S}= -\frac{3}{4}J$ and $ E_{T}=\frac{J}{4}$ respectively. 

For the singlet, the wave functions is $\ket{S}=\frac{1}{\sqrt{2}}(\ket{\downuparrows}-\ket{\updownarrows}) $ and for the triplet, the eigenvectors are $\ket{T_{+}} =  \ket{\upuparrows}$, $ \ket{T_{-}}=\ket{\downdownarrows}$, $\ket{T_{0}} =  \frac{1}{\sqrt{2}}(\ket{\downuparrows}+\ket{\updownarrows})$.
Now, we calculate the different terms in the spin susceptibility formula. The spin correlator, $\langle S^{\alpha}_i S^{\alpha}_j\rangle$, can be divided when $i=j$ and $i\neq j$. We start with $i=j$, i.e., the expected value of ${S_i^{\alpha}}^2$:

\begin{equation}
    \langle {S_i^{\alpha}}^2 \rangle = \frac{ \sum_n \langle n| {S_i^{\alpha}}^2|n \rangle e^{-\beta E_n}}{\sum_n e^{-\beta E_n}}.
\end{equation}
For any spin, the value of  $\langle n| {S_i^{\alpha}}^2|n \rangle = \frac{1}{4}$, therefore, $\langle {S_i^{\alpha}}^2 \rangle =\frac{1}{4} $.
For the case $i\neq j$, we have the expected value of the correlators $\langle S_1^{\alpha} S_2^{\alpha}\rangle$ and $\langle S_2^{\alpha} S_1^{\alpha}\rangle$. Evaluating the wavefunctions for the expected value of the correlator leads to:
\begin{equation}
\begin{split}
     \langle S| S_i^{\alpha} S_j^{\alpha}|S \rangle =&\langle T_0| S_i^{\alpha} S_j^{\alpha}|T_0 \rangle =  -\frac{1}{4}, \\
    \langle T_-| S_i^{\alpha} S_j^{\alpha}|T_- \rangle = & \langle T_+| S_i^{\alpha} S_j^{\alpha}|T_+ \rangle = \frac{1}{4}.
\end{split}
\end{equation}
Therefore:
\begin{equation}
    \langle S_i^{\alpha} S_j^{\alpha}\rangle = \frac{ \sum_n \langle n| S_i^{\alpha} S_j^{\alpha}|n \rangle e^{-\beta E_n}}{\sum_n e^{-\beta E_n}}=\frac{1}{4} \frac{e^{-\beta J/4 }-e^{\beta J3/4}}{3e^{-\beta J/4 }+e^{\beta J3/4 }} ,
\end{equation}
where $\langle S_1^{\alpha} S_2^{\alpha}\rangle =\langle S_2^{\alpha} S_1^{\alpha}\rangle$.

Finally, we obtain the square of the expected value of each spin, $\langle  S_i^{\alpha}\rangle^2$:

\begin{equation}
\begin{split}
   \langle S| S_i^{\alpha}|S \rangle  = & \langle T_0| S_i^{\alpha}|T_0 \rangle = 0, \\
    \langle T_-| {S_i^{\alpha}}^2|T_- \rangle =& -\frac{1}{2} \quad \langle T_+| {S_i^{\alpha}}^2|T_+ \rangle = \frac{1}{2}.
\end{split}
\end{equation}

The energies of $\ket{T_-}$ and $\ket{T_+}$ are the same for no external magnetic field applied, therefore the terms cancel each other and $\langle S_i^{\alpha} \rangle =0$.

Substituting all the terms on Eq. (\ref{chi}), we obtain:

\begin{equation}
    \chi_{ \alpha}= (\mu_B g)^2 \beta[2 \langle S_1^{\alpha}S_2^{\alpha} \rangle + 2 \langle {S_1^{\alpha}}^2 \rangle ]  = (\mu_B g)^2 \beta \frac{2}{3+e^{\beta J }}.
\end{equation}

Now, we obtain the limit at $T \rightarrow 0$  and $T \rightarrow \infty$ for $J>0$:

\begin{equation}
\begin{split}
    \lim_{T \to 0} \chi_{ \alpha} (T,J)= & \lim_{\beta \to \infty}  (\mu_B g)^2 \beta \frac{2}{3+e^{\beta J }} =  \lim_{\beta \to \infty} \frac{2 (\mu_B g)^2 }{J e^{\beta J }} = 0,\\
    \lim_{T \to \infty} \chi_{ \alpha} (T,J)= & \lim_{\beta \to 0}  (\mu_B g)^2 \beta \frac{2}{3+e^{\beta J }} = \beta\frac{ (\mu_B g)^2  }{2}.
\end{split}
\end{equation}

At $T \rightarrow 0$, only the $S=0$ is occupied and the spin susceptibility goes to zero.  In the opposite limit, for large temperatures, the leading term scales with $T^{-1}$  but lies in the region where entanglement can not be certified.

\section{Dipolar Interaction}

In general, surface spins are not isotropic and therefore their susceptibilities do not satisfy eq. (3). In this regard, $S=1/2$ are very convenient as they do not have single-ion anisotropy. Still, magnetic dipolar interactions between the spins in the array are certainly present. Spin arrays are usually fabricated so that the dominant spin-rotationally invariant exchange prevails. Therefore, we can expect the dipolar interaction will lead to a small correction to the spin susceptibility. We thus write the spin susceptibility,  measured experimentally, as:
\begin{equation}
\overline{\chi}_{\alpha}= \beta \left( \sum_{i,j}\langle m_{\alpha}^i m_{\alpha}^j\rangle -(\sum_i \langle m_{\alpha}^i\rangle )^2\right)  +\delta \chi_{\alpha}^{\rm dip}.
\end{equation}
Since we determine experimentally $\overline{\chi}_{\alpha}$
and we can compute $\delta \chi_{\rm dip}^\alpha$, we can now write the sufficient condition to certify entanglement as:
\begin{equation}
\sum_{\alpha}\overline{\chi}_{\alpha}- \sum_{\alpha}\delta \chi_{\rm dip}^\alpha 
\leq \frac{g_{\rm min}^2\mu_B^2S}{k_BT}N.
\label{EW2}
\end{equation}
Now, if $\sum_{\alpha}\delta \chi_{\rm dip}^\alpha$ is positive, then the observed susceptibility is an upper limit for the left hand side of eq. (4), and it is enough to verify that the observed susceptibility satisfies eq. (4) to certify the presence of entanglement: 
\begin{equation}
 \sum_{\alpha}\left(\overline{\chi}_{\alpha}- \delta \chi_{\alpha}{\rm dip}\right)
 \leq 
\sum_{\alpha}\overline{\chi}_\alpha
\leq
 \frac{g_{\rm min}^2\mu_B^2}{k_BT}S .
\label{EW3}
\end{equation}
On the contrary, if $\sum_{\alpha}\delta \chi_{\rm dip}^\alpha$ is negative, we need to resort to theory to make sure that the difference between the entanglement boundary and the observed susceptibility is larger than the calculated  dipolar contribution. For the $N=3$ chains we have found that $\sum_{\alpha}\delta \chi_{\rm dip}^\alpha<0$ but its relative contribution is below 1 percent even for modest values of $J\simeq 1$ meV (see suppl. mat.).  We note that even in the case when dipolar interactions are dominant, a heuristic relation between entanglement and spin susceptibility has been demonstrated \cite{ghosh03}.

The dipolar interaction enters the spin Hamiltonian of the chains as:
\begin{equation}
    {\cal H}_{\rm dip}= \frac{\mu_0}{4\pi} \sum_{i,j} \frac{\vec{m}_i\cdot \vec{m}_j}{r_{ij}^3} - 3 \frac{(\vec{m}_i\cdot\vec{r}_{ij})( \vec{m}_j\cdot\vec{r}_{ij})}{r_{ij}^5}.
\end{equation}
We thus introduce the dipolar energy scale:
\begin{equation}
    \epsilon_{\rm dip}\equiv \frac{\mu_0 g^2 \mu_B^2}{a^3}
\end{equation}
where $a$ is the first-neighbour distance of the spin chains considered here.  The ratio between exchange $J$ and $\epsilon_0$ permits to anticipate the relative importance of dipolar couplings. 
For $a=2.97\AA$, the oxygen-oxygen distance on the surface of MgO, we have $\epsilon_{\rm dip}=8.13\mu$eV, for $g=2$. Thus, for $J=0.1$ meV, a rather modest value of exchange, we have $J>12\epsilon_{dip}$.

\begin{figure}[H]
\centering
\includegraphics[width=0.7\linewidth]{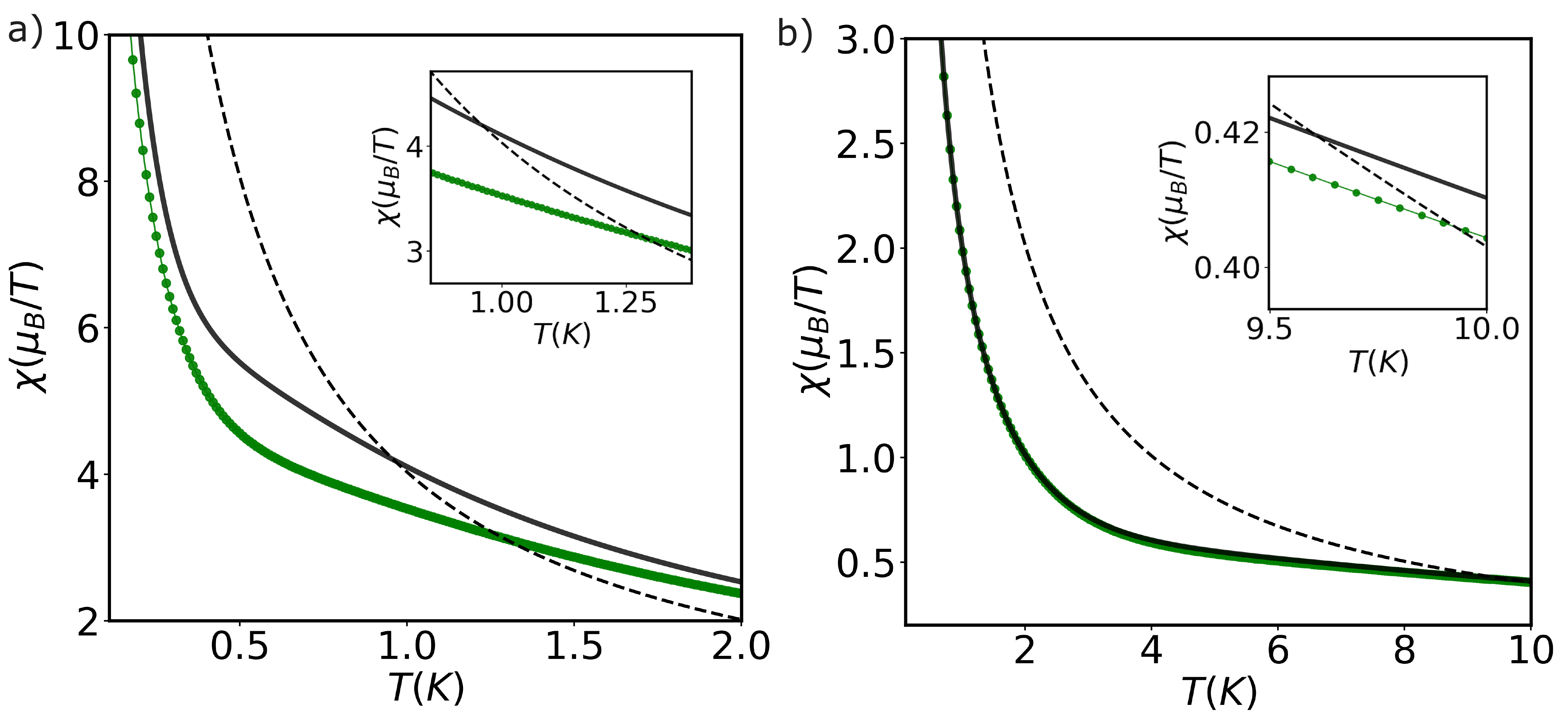}
\caption{Comparison of the spin susceptibility  with (green dots) and without (solid black line) dipolar interactions. Entanglement is certified in the region under  the dashed line. a) $J = 10 \epsilon_{dip}$ and b)  $J = 100 \epsilon_{dip}$. Inset, a zoom near $T \approx T^*$.}
\label{fig:Dipolar}
\end{figure}

In Figure \ref{fig:Dipolar}, we compare the spin susceptibility of an AF spin trimer with and without dipolar interaction for two values of $J/\epsilon_0$ for a trimer. In Figure \ref{fig:Dipolar} a, we show the case of  $J=10\epsilon_{dip}=81.3\mu$eV.  The correction of the dipolar interaction (green dots) is in the order of 10 percent for the $T$ close to $T^*$.  In contrast, for  $J=100\epsilon_{dip}=0.81$ meV (Figure \ref{fig:Dipolar} b ), still a modest energy, the  correction introduced by the dipolar interaction is smaller than one percent.


\end{appendix}

\bibliography{biblio}